\documentclass[showpacs,aps,prl,twocolumn,floatfix,superscriptaddress]{revtex4}
\usepackage{bbding}
\usepackage{epsfig}
\usepackage{color}
\usepackage{graphicx}
\usepackage{dcolumn}
\usepackage{bm}
\usepackage{ulem}
\usepackage{lipsum}
\usepackage{amssymb}
\usepackage{amsmath}
\newcommand{\be}{\begin{equation}}
\newcommand{\ee}{\end{equation}}
\newcommand{\bea}{\begin{eqnarray}}
\newcommand{\eea}{\end{eqnarray}}

\newcommand{\hbbar}{}
\newcommand{\sgn}{{\rm sign}}
 
\begin{document}
\title{A quantum resonance catastrophe for transport through an AC driven impurity} 

\author{Daniel Thuberg}
\affiliation{Instituto de F\'isica, Pontificia Universidad Cat\'olica de Chile, Casilla 306, Santiago 22, Chile}
\author{Sebasti\'an A. Reyes}
\affiliation{Instituto de F\'isica, Pontificia Universidad Cat\'olica de Chile, Casilla 306, Santiago 22, Chile}
\author{Sebastian Eggert}
\affiliation{Physics Department and Research Center OPTIMAS, University of Kaiserslautern, D-67663 Kaiserslautern, Germany}

\begin{abstract}

We consider the quantum transport in a tight-binding chain with a 
locally applied potential which is oscillating in time.
The steady state for such a driven impurity can be calculated exactly for any 
energy and applied potential using the Floquet formalism.  The resulting transmission
has a non-trivial, non-monotonic behavior depending on incoming momentum, 
driving frequency, and the strength of the applied periodic potential. Hence there is 
an abundance of tuning possibilities, which allows to find resonances of total 
reflection for any choice of incoming momentum and periodic potential.
Remarkably, this implies that  even for an arbitrarily small
infinitesimal impurity potential it is always possible to find a 
resonance frequency at which there
is a catastrophic breakdown of the transmission $T=0$.
The points of zero transmission are closely related to the phenomenon of Fano resonances
at dynamically created bound states in the continuum.  
The results are relevant for a variety of one-dimensional systems where local 
AC driving is possible, such as quantum nanodot arrays, ultracold gases in optical 
lattices, photonic crystals, or molecular electronics.
\end{abstract}

\maketitle

%\noindent{\it Introduction}.---
Driven quantum systems 
appear in many different contexts in physics and
chemistry \cite{book,grifoni,van2002electron,della2007visualization,
driven-plasmon,oberthaler,dressed-matter,harper,shaking1,chern-numbers}.
At the same time there has been remarkable progress in  
the controlled design of nanoscale quantum systems with a high degree 
of coherence and tunability. Systems containing just a few molecules are 
promising candidates for the realization of 
electronic components on the sub-silicon 
scale (molecular electronics) \cite{balzani2006molecular, nitzan2003electron, reed1997conductance, cui2001reproducible, aviram1974molecular,goser2004nanoelectronics}. 
The quantum transfer of particles between such localized
structures can be well described by the tight-binding model in order to 
gain 
understanding of the transport mechanisms 
involved \cite{fagas2001electron, cuniberti2002fingerprints,shen2010electron, gutierrez2002theory}. 
Another versatile realization of near-perfect tight-binding models are ultracold gases 
in optical lattices with a great variety of possible geometries \cite{bloch},
where tunable local impurities \cite{widera}, 
periodic driving \cite{chern-numbers,oberthaler,dressed-matter,harper,shaking1}, and
dimensional crossover \cite{ott} have also been realized.
Tight-binding 
models are also applicable when investigating other driven systems such as 
quantum dot arrays \cite{van2002electron}.  Finally, photonic cavities 
and  photonic 
crystals have been used as quantum simulators to fabricate interesting
quantum tight-binding systems \cite{photonic-floquet,photonic,photonic2}. 
In practical applications time-dependent effects such as  
electromagnetic radiation or gate voltages can be used to manipulate the 
transport properties of nanodevices \cite{van2002electron}. 
Driven tunneling between two
 quantum wells has been 
well studied \cite{van2002electron,della2007visualization,driven-plasmon,oberthaler}.
A natural further
development is the transport through a driven impurity in an extended structure of coupled
wells with a finite bandwidth.

In the present Letter we consider the steady state of a generic model system, 
consisting of a one-dimensional tight-binding chain for bosons or fermions
with a 
periodically varying potential $\mu$ at one impurity site ($i=0$)
\be
H = -J \sum_i (c_i^\dagger c_{i+1}^{\phantom{\dagger}} \!\! + 
c_{i+1}^\dagger c_{i}^{\phantom{\dagger}})  -  \mu \cos (\omega t) c_0^\dagger c_{0}^{\phantom{\dagger}},
\label{model}
\ee
where we have used standard notation and the hopping amplitude is denoted by $J$.
This model captures the essential physics of a single one-dimensional band, which is
useful for the description of 
corresponding AC driven experimental setups mentioned above.  Higher energy bands may give
interesting additional effects at very high frequencies, but this will not 
change the central results in this work
since the most interesting physics turns out to involve excitations close to the band-edge.
%as we will see below.
%In fact, it turns out that the calculations can be generalized
%to a {\it continous} model of particles moving inside a wire with a local
%time-periodic potential.

%As we will show below,
%the exact transmission coefficient through such a driven impurity potential 
%can be obtained using 
%the Floquet formalism, which allows a treatment beyond the perturbative regime. 
%The resulting transport properties display a highly non-trivial behavior,  which 
%depends non-monotonically on the driving strength $\mu$, the driving frequency $\omega$, 
%and the incoming momentum $k$.   
It is well-known that for a static barrier, 
the phenomenon of tunneling allows transport for any finite potential strength $\mu$.
In strong contrast, the time-periodic potential considered here turns out to show
resonances at special driving frequencies where the transmission is completely blocked.
In fact, we will show that 
it is always possible to find such a finite resonance frequency $\omega$
for any combination
of incoming momentum $k$ and barrier strength $\mu$.  
This implies that even for an {\it 
infinitesimally small} periodic perturbation $\mu$  there is a complete breakdown
of conductance, if the frequency $\omega$ is 
tuned to the corresponding resonance.  We call this
phenomenon the {\it quantum resonance catastrophe}.

\begin{figure}[!t]
\begin{center}
\includegraphics[width=0.37\columnwidth]{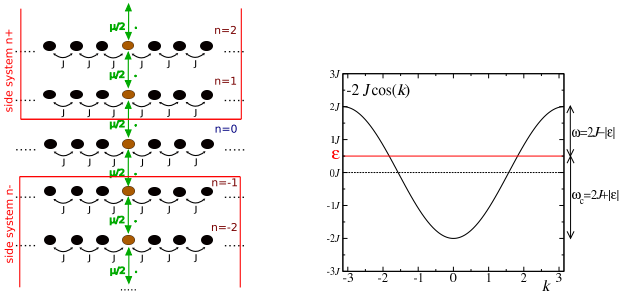} ~~~~~~~~
\includegraphics[width=0.45\columnwidth]{disp}
\end{center}
\caption{{\it Left:} Sketch of the model mapped onto a set of static coupled chains. 
The $n=0$ chain is locally connected to two side-coupled systems of chains with 
a corresponding chemical potential $n\hbbar \omega $.
{\it Right:} Dispersion relation with the special frequencies 
$\hbbar \omega = 2J \pm \epsilon$ below which the 
side coupled chains $n=\mp 1$ start 
support unbound solutions 
for the case $\epsilon=0.5J$. }
\label{chains}
\end{figure}

The goal of this paper is to calculate the transport of an incoming particle 
with a given momentum $k$ and corresponding energy $\epsilon =-2 J \cos k$, which 
is the dispersion relation of the model in Eq.~(\ref{model}) away from the impurity.  
Just like in the static case, 
the transmission coefficient can be determined from
the steady state solution of the Schr\"odinger equation
$(H(t)-i\hbbar \partial_t)|\Psi(t)\rangle=0$.  
%{\color{red} It is also possible to consider
%arbitrary incoming wave-packets by using a superposition of those solutions.}
Due to the periodicity of the 
Hamiltonian $H(t) = H(t+2\pi/\omega)$ it is possible to use the Floquet 
formalism \cite{floquet,grifoni}
to express any steady state solution in terms
of so-called Floquet states $\left|\Psi(t)\right> = 
e^{-i \epsilon t/\hbbar}\left|\Phi(t)\right>$, 
where $\epsilon$ is the quasi-energy of the resulting 
(d+1)-dimensional eigenvalue equation 
$(H -i\hbbar \partial_t) \left|\Phi(t)\right> = \epsilon \left|\Phi(t)\right>$ and
the Floquet modes
 $\left|\Phi(t)\right> = \left|\Phi(t+2 \pi/\omega)\right>$ are periodic
in time.  Using the spectral decomposition
\be
|\Phi(t)\rangle = \sum_{n=-\infty}^{\infty}e^{-i n \omega t} |\Phi_{n}\rangle, \label{Eq:spectral}
\ee
the eigenvalue equation for a Hamiltonian of the form {$H(t) = H_0 +2 H_1 \cos(\omega t)$}
becomes discrete in the time direction
\be
H_0 |\Phi_{n}\rangle + H_1 (|\Phi_{n+1}\rangle + |\Phi_{n-1}\rangle) = 
(\epsilon + n \omega) |\Phi_{n}\rangle.
\ee
A general steady state on the tight-binding lattice is given by
\be
|\Phi_{n}\rangle = \sum_j \phi_{j,n} c_j^\dagger |0\rangle. \label{Eq:staten}
\ee
The model in Eq.~(\ref{model}) therefore results in the following set of coupled equations
\bea
& &\!\!\!\!\!\!\!\! -J (\phi_{-1,n}+\phi_{1,n})
-\frac{\mu}{2}(\phi_{0,n+1}+\phi_{0,n-1})  
=  (\epsilon+n\hbbar\omega) \phi_{0,n}  
\nonumber \\
& &\!\!\!\!\!\!\!\! -J (\phi_{j-1,n}+\phi_{j+1,n})  =  (\epsilon+n\hbbar\omega) \phi_{j,n} 
 {\rm \ \ \ for \ \ \  }  j \neq 0 \ 
\label{condnj} 
\eea
which effectively corresponds to a static Hamiltonian with eigenvalue $\epsilon$ for
an infinite number of chains labeled by $n$, each with additional overall 
chemical potential of  
$n \hbbar\omega$, analogous to a Wannier-Stark ladder \cite{grifoni}. 
The chains are coupled to each other only at site $j=0$ %with their nearest neighbor 
by a hopping term $\mu/2$ as depicted in Fig.~\ref{chains} (left). 
Notice that the entire problem is symmetric under parity transformation $j\to -j$, 
so that solutions are either parity symmetric or parity antisymmetric.
The parity anti-symmetric solutions obey $\phi_{0,n} = 0, \  \forall n$, so they do not 
couple to the driving potential and can be ignored.

\noindent{\it Transmission coefficient}.---
We now want to calculate the transmission of an incoming particle with 
momentum $k$ and $\epsilon = -2J \cos k$ for the chain $n=0$.  
The parity symmetric solution is given by 
plane waves of the general form
\be
\phi_{j, 0} = A \cos(|j| k - \theta).
%\frac{C_0}{2 \cos \theta}
\label{unbound}
\ee
Since the potential $\mu$ only affects a single impurity site for all chains, 
the solutions for $j\neq 0$
must correspond to wave-like states (unbound solutions) for $|\epsilon + n \omega| <2 J$ and
bound states otherwise according to Eq.~(\ref{condnj}).   
As indicated in Fig.~\ref{chains} (right) a
critical frequency can be defined
\be
\hbbar \omega_c \equiv  2J +|\epsilon|.
\label{omegac}
\ee
For $\omega > \omega_c$ 
all chains with $n\neq 0$ are outside the band
$|\epsilon + \hbbar n \omega|\! >\! 2J$ and correspond to bound states.
Below the frequency $\omega\!=\!\omega_c$ the 
first side coupled chain starts to support unbound solutions.
A second unbound solution starts to appear below $\hbbar \omega\! =\! 2J -|\epsilon|$
and so on.

Let us first consider frequencies $\omega > \omega_c$ with
bound states for all  $n\neq 0$ of the form 
\be 
{%\color{red} 
\phi_{j,n} = C_n e^{- \kappa_n |j|} \sgn(-n)^{j+n}}, 
\label{bound} 
\ee
where
$\epsilon +n\hbbar \omega =2 J \sgn(n) \cosh \kappa_n$.
Inserting these states into Eq.~(\ref{condnj}), 
we arrive at a recurrence relation
for the coefficients $C_n$ for $|n| > 0$,
\be 
\gamma_n C_n= C_{n-1}+C_{n+1} \ {\rm with}\  \gamma_n = 
\frac{2}{\mu} \sqrt{(\epsilon+n \hbbar\omega)^2 - 4 J^2} ,
\label{C} 
\ee
where we have defined $C_0 \equiv A \cos \theta$.
The solution for this second order recurrance relation is fixed up to an overall constant
by requiring convergence for $|n| \to \infty$ and can be solved efficiently numerically.
The angle $\theta$ is then given by 
Eq.~(\ref{condnj}) for $n=0$  in terms
of those coefficients 
\be
{%\color{red} 
\tan\theta = \frac{\mu}{2 u_k}\left(\frac{C_{-1}}{C_0} - \frac{C_1}{C_0}\right)\label{tantheta}
}
\ee
where we defined $u_k\equiv 2J\sin k$ as the particle velocity. 

Since the bound states for $|n|>0$ do not contribute in the transmission, it is 
now straight-forward to calculate the transmission coefficient to be
\be 
{%\color{red} 
T = \cos^2 \theta = 
\frac{u_k^2}{u_k^2 + (\mu(C_{-1} - C_1)/2 C_0)^2}.  \label{Eq:Transmission}
}
\ee
The transmission obeys $T(\epsilon)=T(-\epsilon)$.
For $\epsilon = 0$ the solution becomes symmetric $C_n=C_{-n}$, which results in 
$T(\epsilon= 0) = 1$ independent of $\mu$ {for $\omega > \omega_c$ due to Klein
tunneling \cite{klein}. 	
In the following we assume $\epsilon \neq 0$.}

Next we consider lower frequencies $\omega < \omega_c$, when unbound states also 
exist for $n\neq 0$.  In this case it is useful to 
make an ansatz for an incoming wave at energy $\epsilon$ and transmitted/reflected
waves in all unbound channels \cite{energy}.
In this way it is possible to solve for all parameters. Finally, the
total 
transmission coefficient can again be expressed 
by the solution of the recurrence relation in Eq.~(\ref{C}),
which now involves the remaining bound states.

\begin{figure}[!t]
\includegraphics[width=0.99\columnwidth]{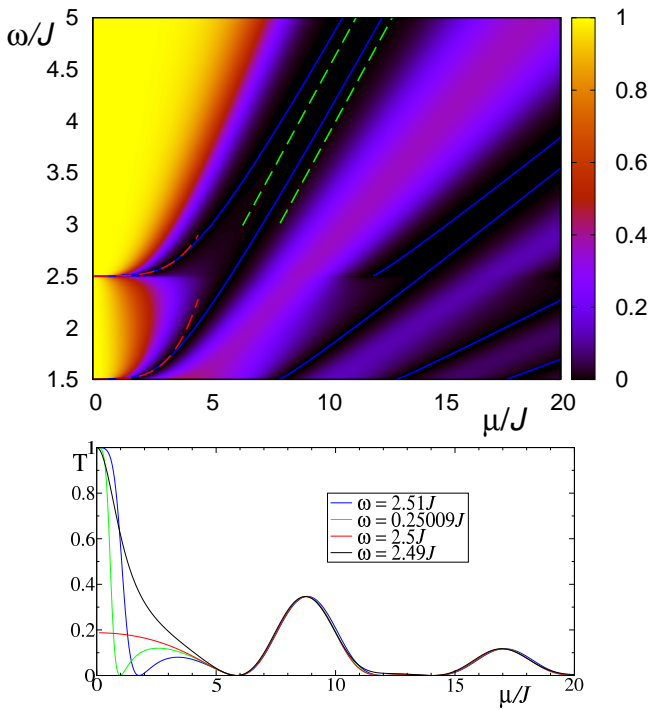}
\includegraphics[width=0.79\columnwidth]{Tvsmu.eps}
\caption{Exact results for transmission coefficient $T$ as a function
 of amplitude $\mu$ and frequency $\omega$ for an incoming
wave of energy $\epsilon = 0.5J$. {\it Top:} The solid lines (blue) indicate the exact values 
of the resonances ($T=0$)  while the dashed lines are analytical expressions 
from the first zeros of Bessel functions 
$J_{\pm \epsilon/\hbbar \omega}$ in Eq.~(\ref{besselapprox})  at large frequencies 
and the small $\mu$ approximations in Eq.~(\ref{smallmu}). {\it Bottom:}
Behavior close to the critical 
frequency $\hbbar \omega \approx \hbbar \omega_c = 2.5J$. For $\omega = \omega_c$ the
transmission approaches $T\to (2J-|\epsilon|)/8J$ as $\mu\to 0$.}\label{fig:T}
\end{figure}

\noindent{\it Results}.--- 
Using this procedure the 
exact numerical solution for the transmission coefficient was obtained as 
shown in 
Fig.~\ref{fig:T} for a given energy $\epsilon = 0.5J$ as a function of $\mu$ and $\omega$.
Perfect transmission $T\to 1$ can be observed for small $\mu$ or large frequencies.
For increasing $\mu$ there is a sharp drop, however, and at special resonances
a vanishing transmission 
$T=0$ can be observed (blue solid lines). % which will be discussed below.
%The data shows sharp gradients in the transmission coefficient close to 
%\be
%\hbbar \omega_c = 2J +|\epsilon|,
%\label{omegac}
%\ee
%which defines the critical frequency below which the first side-coupled chain 
%supports unbound solutions as indicated in Fig.~\ref{chains} (right).
%The next unbound state appears at $\hbbar \omega = 2J - |\epsilon|$.
%These resonances approach $\mu\to 0$ at the values $\hbbar \omega = 2J +|\epsilon|$
%and $\hbbar \omega = 2J - |\epsilon|$, which are exactly the frequencies below which the 
%$n=-1$ and the $n=1$ chains start to support unbound solutions, respectively.
Interestingly, the transmission then {increases} again with 
{increasing} $\mu$ before more resonances with $T=0$
are reached  and so on.  This apparent non-monotonic behavior with 
potential $\mu$ and frequency $\omega$ can be understood in 
the high frequency limit, where the recurrence relation can be solved analytically.
In particular, note that 
$\gamma_n \to 2 |n \hbbar \omega + \epsilon|/\mu$ 
for $\hbbar \omega \gg J$, so that Eq.~(\ref{C}) becomes 
exactly the defining recurrence relation for the Bessel functions \cite{handbook}
in this limit.
{For convergence as $|n| \to \infty$ the coefficients therefore can be chosen
to be Bessel function of the first kind
\be
C_{n} \approx J_{|n+\epsilon/\hbbar \omega|} (\mu/\hbbar \omega).
%\ {\rm and} \ C_{n<0} \propto J_{-n+\epsilon/\hbbar \omega}  (\mu/\hbbar \omega),
\ee
The recurrence relation then approaches $C_0 \approx J_{\pm \epsilon/\omega}$ 
for $n\to 0^\pm$ from above/below.
Accordingly Eq.~(\ref{tantheta}) can be approximated for $\hbbar \omega\! \gg\! J$
\be \tan \theta \approx  \frac{\mu}{2 u_k}\left(
\frac{J_{1-\epsilon/\hbbar \omega}(\mu/\hbbar \omega)}{J_{-\epsilon/\hbbar \omega}(\mu/\hbbar \omega)} - 
\frac{J_{1+\epsilon/\hbbar \omega}(\mu/\hbbar \omega)}{J_{\epsilon/\hbbar \omega}(\mu/\hbbar \omega)} \right) . \label{besselapprox}
\ee}
These  Bessel functions explain some of the 
observed features of $T$ in Eq.~(\ref{Eq:Transmission}), namely the oscillating behavior with $\mu$ and $\omega$
and resonances of $T=0$ close to 
  the zeros of the Bessel functions
$J_{\pm \epsilon/\hbbar \omega}$, which are marked in Fig.~\ref{fig:T} as dashed lines (green).
Moreover, the behavior in Fig.~\ref{fig:T} indeed
only depends on the ratio $\mu/\omega$ for large frequencies.
%It must be mentioned that the zeroth Bessel function $J_0$ has been obtained 
%as a high-frequency approximation
%in several time-periodic problems using the Floquet theory 
%\cite{della2007visualization,shaking}, but the observation
%of Bessel functions with a fractional index $\nu = \pm \epsilon/\hbbar \omega$ has not
%been reported to our knowledge. 

While the description in terms of Bessel functions is useful, this does not explain
the behavior in the most interesting region close to 
$\omega \approx \omega_c$ where the data
% in Fig.~\ref{fig:T} 
shows large gradients
in the transmission coefficient.
In Fig.~\ref{fig:T} (bottom) 
the behavior changes dramatically with frequencies
just above or below $\hbbar \omega_c \!= \!2.5J$
for small driving potential $\mu$. 
There is a resonance with $T=0$ which
quickly shifts to smaller $\mu$ as the frequency is 
lowered ($\hbbar \omega \!= \!2.51J$ and $2.5009J$) and suddenly disappears
completely once the $n\!=\!-1$ 
chain supports unbound solutions ($\hbbar \omega\! =\! 2.49 J$).
Exactly at the critical frequency $\hbbar \omega_c \!= \!2.5J$ 
%the resonance at $\mu\to 0$ cannot be resolved and 
the results for $\mu\to 0$ show a well-behaved
finite value $T\to (2J-|\epsilon|)/8J$, which 
corresponds to $C_{-1}^2/C_0^2 \to  \gamma_{-2}^2$
and is neither close to unity nor zero.
By looking very carefully one observes that there is another resonance 
close to $\mu\approx 11.5J$ which disappears
at $\hbbar \omega_c$. 
Away from these singular points the changes of the transmission are very small, however,
and appear to be continuous as the frequency goes through the critical value.

\noindent{\it Resonances}.--- 
%We now focus on the locations of the resonances with $T=0$ close to the 
%critical frequency.  
It is worth noticing that the resonances $T=\cos^2\theta =0$
for $\omega>\omega_c$ 
are special 
points at which the coefficient $\phi_{0,0} = C_0 = A \cos{\theta}$ vanishes
exactly.  In this case 
the side coupled systems of the corresponding static problem in Eq.~(\ref{condnj}) 
become decoupled from the chain with $n=0$ in 
Fig.~\ref{chains} (left).
{\it Therefore a resonance with $T=0$ for $\omega>\omega_c$
occurs if and only if the isolated side system has an 
eigenenergy inside the band  $|\epsilon| < 2 J$.}

This is a remarkable statement since the decoupled side chains for $n\neq 0$
only support bound states
outside the band.  {However, due to the local coupling $\mu$ between the chains
one of these energies is pushed inside the band, for which 
 $T(\epsilon) =0$.}
Such {\it Bound States in the 
Continuum} (BIC) were first proposed by von Neumann and Wigner 
for a {\it spatially} oscillating potential in the
early days of quantum mechanics \cite{wigner}.
Since then BIC's have received extensive attention in the context of transport 
phenomena \cite{BIC,gonzalez2013bound,longhi2013floquet}.  The suppression of 
transmission is closely related to the Fano-Effect in this case \cite{fano,miroshnichenko2005engineering}.
%In fact, it has been shown that static
%side coupled systems are a tool to engineer the Fano
%resonances \cite{miroshnichenko2005engineering}. 
%As depicted in Fig.~\ref{chains} (left) the setup of a local
%AC driving now allows 
%to create and tune such 
%side coupled systems dynamically.

\begin{figure}[!t]
\begin{center}
\includegraphics[width=0.82\columnwidth]{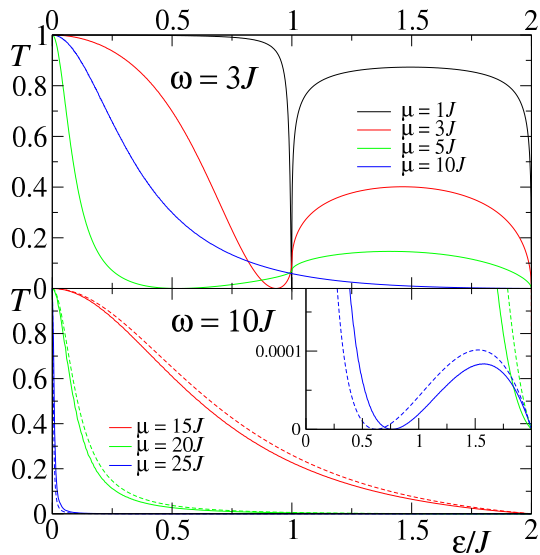}
 \end{center}
\caption{Transmission coefficient as a function of incoming particle energy $\epsilon$
at $\hbbar\omega=3J$ (top) and $\hbbar\omega=10J$ (bottom). 
The dashed lines in the bottom plot depict the high frequency approximation in 
Eq.~(\ref{besselapprox}). The inset shows an enlarged region of 
the resonance for $\mu = 25J$.}
 \label{fano}
\end{figure}

In order to illustrate the connection with the Fano effect the 
behavior as a function of incoming energy at fixed frequency $\hbbar \omega =3J$
is shown in Fig.~\ref{fano} (top).  The characteristic asymmetric lineshape of Fano resonances
is clearly visible.  In analogy to the critical frequency in Eq.~(\ref{omegac}), 
it is possible
to define a critical energy $\epsilon_c= \hbbar \omega -2 J$, above which the
first side chain supports an unbound solution.  We observe that a sharp
resonance occurs just below this energy for small $\mu$ which then
broadens and moves quickly away from this point with increasing $\mu$.  
The lower part of Fig.~\ref{fano} shows 
the behavior for larger frequency $\hbbar \omega = 10J$, where the resonances
are further apart and less pronounced (see inset).
In this limit the agreement with the high frequency approximation in Eq.~(\ref{besselapprox})
fits reasonably well (dashed lines in Fig.~\ref{fano}).
The drop of the transmission $T\to 0$ for $\epsilon \to  2J$ occurs due to the
vanishing of the particle velocity $u_k \to 0$ and is not connected to any 
resonance phenomenon.  

As shown in Fig.~\ref{fano} for fixed $\mu$
and $\omega$ there is at most one energy with $T=0$ and
in some cases it is not possible to find a resonance at all, 
e.g.~for $\hbbar \omega=3J$ and $\mu=10J$. 
This is in contrast to the situation of a given energy discussed above in 
Fig.~\ref{fig:T} where there are always one or more  resonance frequencies
for any value of $\mu$. 
%(and several values of $\mu$ for a given frequency).

In order to predict the location of the resonances with $T=0$ we
now follow the strategy 
to consider 
the eigenenergies of the decoupled side system in 
Fig.~\ref{chains}
using Eq.~(\ref{condnj}) with $\phi_{0,0}=0$.
In fact, the two sides for 
$n\!>\!0$ and $n\!<\!0$ have identical 
eigenenergies $\epsilon$, due to the symmetry transformation
$\phi_{j,n} \to (-1)^{n+j} \phi_{j,-n}$.
Of course there are infinitely many eigenenergies, but only those inside the
band $|\epsilon| < 2J$ are of interest.
Let us focus on the resonances $T=0$ close to the critical frequency $\omega\approx \omega_c$
in the limit of small $\mu \ll J$.  In this case, Eq.~(\ref{C})
must still hold for $C_0=0$, with $\gamma_{n\neq 1} \gg 1$ and $\gamma_1 \ll 1$.
Looking at the first few terms of the recurrence relation it becomes clear that
the coefficients grow beyond bounds unless $1-\gamma_1 \gamma_2 \ll \gamma_1 \ll 1$. 
Solving for the frequency at which $\gamma_1 \gamma_2 = 1$ we find
for the resonance positions
\be
\hbbar\omega\approx \hbbar \omega_c +\frac{\mu^4}{64 J\left([4 J-\epsilon]^2-4J^2\right)},
\label{smallmu}
\ee
which is marked by dashed lines (red) in Fig.~\ref{fig:T} and agrees well with the
exact results.
Using the condition $\gamma_1 \gamma_2 = 1$, resonances can also 
be found at fixed $\omega$ for small deviations from the critical 
energy $\epsilon_c$, which are again proportional to $\mu^4$.

\noindent {\it Conclusions}.---
In summary we have developed a framework to study the transmission across
an AC driven barrier connected to leads of finite bandwidth based on the Floquet formalism.
The resulting transmission coefficient $T$ can be calculated exactly and 
shows versatile tunability with frequency $\omega$, energy $\epsilon$, and 
impurity strength $\mu$.
At high frequencies $\omega\gg J$
the transmission can be expressed in terms of Bessel functions, which 
also appear in the description of so-called coherent destruction of tunneling. 
In this limit 
the oscillations can be averaged to form an effective quasi-static
hopping $J_{\mbox{\tiny eff}}=J J_0(\mu/\hbbar\omega)$ \cite{della2007visualization,shaking}.
In our case, a slightly more refined picture emerges in terms of 
Bessel functions with a fractional index $\nu = |n|\pm \epsilon/\hbbar \omega$.

However, much more interesting effects appear at lower frequencies
$\omega \approx \omega_c=2J+|\epsilon|$ where there are sharp changes in $T$
and a complete breakdown of the transmission $T=0$ may appear even for arbitrarily small
barriers $\mu$.  The explanation of such a {\it quantum resonance catastrophe} can be found
in the dynamically created side coupled chains in Fig.~\ref{chains}, 
which contain bound states for all energies
{\it outside} the band for $\omega>\omega_c$. 
{The effect of the local coupling $\mu$ between the chains is to  
push one energy from just above the band  
into the continuum. Thus
effectively  a discrete {\it Bound State in the
Continuum} is formed, which is known to have drastic effects
on the transmission \cite{wigner} based on 
the Fano effect \cite{fano}.} There has been an interesting proposal to use
static side coupled systems to engineer Fano resonances 
\cite{miroshnichenko2005engineering}, but  
the creation of ``virtual'' 
side coupled systems by AC driving considered here is even simpler and more versatile than
any static design.  The location of the resulting resonances for $\mu\to 0$ 
can be predicted rather accurately by Eq.~(\ref{smallmu}).
It should be noted that the width of the resonance
also changes dramatically near $\omega_c$.  
While this may become the limiting factor to observe
the quantum resonance catastrophe for very small $\mu$, it also presents a unique 
opportunity for the design of switches, where a huge change of transmission
for small parameter changes near sharp resonances is a desirable feature.

The underlying tight-binding model has long been 
used in condensed matter, but 
with the recent trend
of designing tailored quantum systems, this model has made a revival
for the realistic description of corresponding setups in 
molecular electronics, quantum nano dots, and photonic materials. 
For ultracold gases in optical lattices \cite{bloch}
it is now possible to insert localized impurities \cite{widera}, which play 
the role of a local barrier that can easily be periodically changed using 
Feshbach resonances.  Given the large variety of applicable 
systems it is difficult to anticipate which experimental 
realization is best suited to explore the quantum resonance catastrophe predicted here.

%Our results apply to independent quantum particles (fermions or bosons), but interactions
%should also be considered in future research.  In one dimension correlations 
%can lead to an interesting additional 
%renormalization of the transmission with energy \cite{kane,kehrein}.

We thank Luis Foa, Pedro Orellana, and Luis Rosales for fruitful discussions. 
%S. R. and D. T. were supported by FONDECYT grant N°11110537. S. E. ...
This research was financially supported by FONDECYT grant No.~11110537, CONICYT grant No. 63140250 and
by the German Research Fundation (DFG) via the SFB/TR49.

 \bibliographystyle{abbrv}
% \bibliography{paper}

\begin{thebibliography}{10}
\bibitem{energy}  Since there is no energy conservation, the particle
may be transmitted/reflected into any of the unbound Floquet modes with $n\neq 0$.
\end{thebibliography}

%Unused bibitems

%\bibitem{tsu1973tunneling}
%R.~Tsu and L.~Esaki,
%\newblock Tunneling in a finite superlattice,
%\newblock {Appl. Phys. Lett.} {\bf 22}, 562 (1973).
%
%\bibitem{tunneling}
%G. Gamow,
%Zur Quantentheorie des Atomkernes,
%Z. Phys. {\bf 51}, 204 (1928).
%\bibitem{CDT}
%F. Grossmann, T. Dittrich, P. Jung, and P. H\"anggi,
%Coherent destruction of tunneling,
%Phys. Rev. Lett. {\bf 67}, 516 (1991).
%
%\bibitem{dot-array}
%C.A. Stafford and S. Das Sarma,
%Collective Coulomb blockade in an array of quantum dots: A Mott-Hubbard approach,
%Phys. Rev. Lett. {\bf 72}, 3590 (1994).

%\end{thebibliography}

\end{document}